\documentclass[prl,floats,twocolumn,showpacs,floatfix,superscriptaddress]{revtex4}
\usepackage{graphics,graphicx,dcolumn,bm,fleqn,epic,eepic,float}
\usepackage{amssymb,amsmath,multirow,rotate,color,float,mciteplus}
\usepackage[latin1]{inputenc}

\begin{document}

\title{Effective Hydrodynamic Boundary Conditions for Microtextured Surfaces}
\author{Anne Mongruel}
\affiliation{Physique et Mécanique des Milieux Hétérogènes (PMMH), UMR 7636 CNRS ; ESPCI
ParisTech ; Univ. Pierre et Marie Curie (UPMC) ; Univ. Paris Diderot (Paris
7) \\
10 rue Vauquelin, 75231 Paris cedex 05, France}
\author{Thibault Chastel}
\affiliation{Physique et Mécanique des Milieux Hétérogènes (PMMH), UMR 7636 CNRS ; ESPCI
ParisTech ; Univ. Pierre et Marie Curie (UPMC) ; Univ. Paris Diderot (Paris
7) \\
10 rue Vauquelin, 75231 Paris cedex 05, France}
\author{Evgeny S. Asmolov }
\affiliation{A.N.~Frumkin Institute of Physical Chemistry and Electrochemistry, Russian
Academy of Sciences, 31 Leninsky Prospect, 119991 Moscow, Russia}
\affiliation{Central Aero-Hydrodynamics Institute, 1 Zhukovsky str., Zhukovsky, Moscow
region, 140180, Russia}
\affiliation{Institute of Mechanics, M. V. Lomonosov Moscow State University, 119992
Moscow, Russia}
\author{Olga I. Vinogradova}
\affiliation{A.N.~Frumkin Institute of Physical Chemistry and Electrochemistry, Russian
Academy of Sciences, 31 Leninsky Prospect, 119991 Moscow, Russia}
\affiliation{Department of Physics, M. V. Lomonosov Moscow State University, 119991
Moscow, Russia}
\affiliation{DWI, RWTH Aachen, Forckenbeckstr. 50, 52056 Aachen, Germany}
\date{\today}

\begin{abstract}
Understanding the influence of topographic heterogeneities on liquid flows has become an important issue with the development of microfluidic systems, and more generally for the manipulation of liquids at the small scale. Most studies of the boundary flow past such surfaces have concerned poorly wetting liquids for which the topography acts to generate superhydrophobic slip.  Here we focus on  topographically patterned but chemically homogeneous surfaces, and measure a drag force on a sphere approaching a plane decorated with lyophilic microscopic grooves. A significant decrease in the force compared with predicted even for a superhydrophobic surface is observed. To quantify the force we use the effective no-slip boundary condition, which is applied at the imaginary smooth homogeneous isotropic surface located at an intermediate position between top and bottom of grooves.
 We relate its location to a surface topology by a simple, but accurate analytical formula. Since groves represent the most anisotropic surface, our conclusions are valid for any texture, and suggest rules for the rational design of topographically patterned surfaces to generate desired drag.

\end{abstract}

\maketitle

\textbf{Introduction.}-- The advent of microfluidics has motivated the growing interest in
understanding and modeling of flows at small scales or in tiny channels. In recent years it has become clear that
the no-slip boundary condition at a solid-liquid
interface is valid only for smooth hydrophilic surfaces~\cite{vinogradova:03,charlaix.e:2005,joly.l:2006,vinogradova.oi:2009}, and for many
other systems it does not apply when the size of a system is reduced. Thus the hydrophobicity of smooth surfaces could
induce a partial slippage, $v_s = b \partial v / \partial z,$ where $v_s$ is
the velocity at the wall, $b$ the slip length, and the axis $z$ is normal to
the surface~\cite{vinogradova.oi:1999}. This concept is
now well supported by nanorheology
measurements~\cite{charlaix.e:2005,vinogradova:03,vinogradova.oi:2009}.

However, only very few solids are molecularly smooth. Most of
them are rough, often at a micrometer scale. This roughness may be induced
by some processes of fabrication or coating, but microtextures are also
found on the surfaces of most plants and animals. In particular, many solids
are naturally striated by grooves, which can also be prepared for
specific microfluidic purposes, such as passive chaotic mixing~\cite{stroock2002a,mixer2010}. Most studies of flow past rough surfaces have
concerned poorly wetting liquids for which the topography acts
to favor the formation of trapped gas bubbles (Cassie state), and to generate superhydrophobic slippage~\cite{Rothstein:2010,vinogradova.oi:2011}. For
rough wettable surfaces the situation is unclear, and opposite
conclusions have been made: one is that roughness generates
extremely large slip~\cite{bonaccurso-03}, and one is that it decreases the
degree of slippage~\cite{zhu-granick-02}. Recent data
(supported by simulations~\cite{kunert.c:2010}) suggest
that the description of flow near rough surfaces has to be corrected, but
rather for a separation, but not
slip~\cite{vinogradova-yakubov-06,steinberger.a:2007}. Another suggestion is to combine these two models~\cite{guriyanova.s:2010}.

\begin{figure}[tbp]
\begin{center}
{\includegraphics[width=7.5 cm]{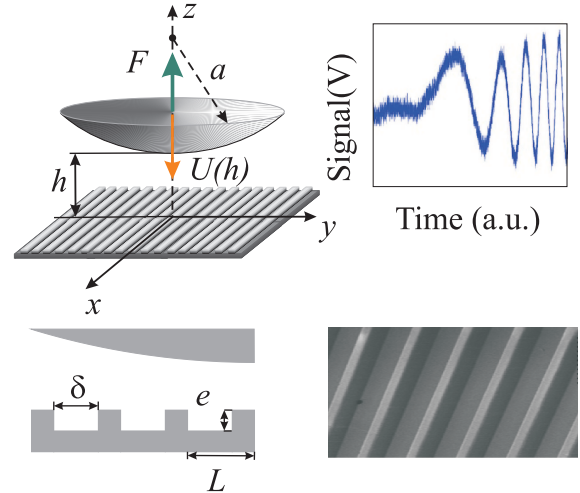}}
\end{center}
\caption{Sketch of a sphere approaching a model grooved surface (left) with
the example of a typical experimental signal and a scanning electron
micrograph (taken under an angle) of the surface obtained by a soft
lithography (right)}
\label{fig:sketch}
\end{figure}

In this Letter we describe how the boundary conditions can be modified by
the surface texture. We focus on the case of special interest where this
model surface is decorated by rectangular microgrooves, i.e. on the situation of the largest possible anisotropy of the flow. We analyze the hydrodynamic interaction between a smooth sphere and a grooved plane, as
sketched in Fig.~\ref{fig:sketch}, and the texture parameters are
systematically varied at the micrometer level, in order to investigate their influence on a drag
force. Our results do not support some previous experimental conclusions on a
 large slip for similar systems. Instead, we unambiguously prove the concept of an effective no-slip plane shifted down from the top of roughness. To the best of our knowledge, this is the first study, where experimentally found values of this shift were quantified theoretically and related analytically to controlled parameters of topographic patterns.

\textbf{Experimental.}-- We use a specially designed homemade
setup~\cite{{Lecoq:1993},{Lecoq:2004},{Mongruel:2010}} to measure on a microscale the
displacement of a sphere towards the corrugated wall at constant gravity
force. The steel sphere of density $\rho _{p}=7.8\times
10^{3}\,$kg\thinspace m$^{-3}$ and radii ranging from $3.5$ mm to $6.35$ mm is
embedded in a liquid contained in a cylindrical glass vessel with a
$50$\thinspace mm diameter and a $40$\thinspace mm height. As a liquid we have chosen high molecular weight
PDMS (silicone) oil (47V100000 Rhodorsyl oil, from Rhone-Poulenc.), with
dynamic viscosity $\mu =97.8$ Pa\thinspace s at $25^{\circ }$C, which is
Newtonian for shear rates up to $100\,$s$^{-1}$. Such shear rates are never reached in our experiment.

\begin{center}
Table 1 : {\small {\textit{Parameters of the textured samples and the shift
of effective hydrodynamic wall, $s$.}} } \begin{tabular}{|c|c|c|c|c|c|c|}
\hline
$N$ & $L$ & $\delta $ & $\phi $ & $e$ & $s$, expe- & $s$, theory \\
&  &  &  &  & riment, &[Eqs.(\ref{shift-SHS}), (\ref{b-para}), (\ref{b-perp})],\\
& ($\mu$m) & ($\mu$m) &  & ($\mu$m) & ($\mu$m) & ($\mu$m) \\ \hline
1 & 100 & 50 & 1/2 & 45 & 5 $\pm$ 0.1 & 5.5 \\ \hline
2 & 150 & 50 & 1/3 & 45 & 4.2 $\pm$ 0.3 & 3.5 \\
3 & 150 & 100 & 2/3 & 45 & 13 $\pm$ 2 & 11.8 \\ \hline
4 & 200 & 100 & 1/2 & 76 & 13 $\pm$ 3 & 10.4 \\
5 & 200 & 100 & 1/2 & 45 & 9 $\pm$ 1.5 & 8.3 \\ \hline
6 & 250 & 25 & 1/10 & 42 & 0.5 $\pm$ 0.1 & 0.6 \\
7 & 250 & 225 & 9/10 & 42 & 28.5 $\pm$ 0.5 & 23.5 \\ \hline
\end{tabular}
\end{center}

The microstructured surfaces were created by common soft lithography, in a
three steps process, transferring geometric shapes from a mask: first to a
silicon wafer coated with a (SU8) photoresist, second to a replica molding
obtained by soft imprint of a thermo-reticulable PDMS, and finally to a
replica of the PDMS mold by soft imprint of a thiolene based resin (NOA 81,
Norland optical adhesives) on glass microscope slides (to be fixed at the
bottom of the vessel). This resin was chosen for its resistance to
compression and to solvent-swelling, and for its good adhesion to glass~\cite{bartolo.d:2008}.
The
final structures are checked by scanning electron microscopy (see Fig.~\ref{fig:sketch}).
The textures are characterized by spacing $\delta $, height $e$ and period
$L $. The liquid fraction, $\phi =\delta /L,$ can be precisely measured since
it is the ratio of the upper surface of the crenellations over the total
surface of the sample. It varies largely with the patterns (from 0.1 to
0.9), and $e/L$ varies from 0.168 to 0.45, as displayed in Table~1. Contact
angles against PDMS for all textures were found to be below $30^{\circ},$ so
that surfaces can be considered as lyophilic. Therefore, we expect PDMS to
invade the surface texture (Wenzel state).

We measure the distance, $h,$ which is defined from the top of the textures
(contact) by using an interferometric technique
\cite{{Lecoq:1993},{Lecoq:2004},{Mongruel:2010}} with the accuracy 0.2 $\mu$m.
The velocity $U(h)$ of the sphere is found by multiplying the velocity of
interference fringes displacement by a factor of $\lambda /2n,$ where
$\lambda =632.8$ nm is the wavelength of the He-Ne laser, and $n=1.404$ the
refraction index of PDMS. After opto-electronic conversion and
amplification, the signal is recorded with a high frequency electronic
oscilloscope (DPO4032 from Tektronics). A deceleration of the sphere (Fig.~\ref{fig:sketch}) is
reflected in the increase of the period of the signal, until contact occurs,
and its position is defined from the recorded signal, at the time when the
period of the signal becomes very large indicating a vanishing velocity.
Note that the signal-to-noise ratio deteriorates at vanishing frequency,
because the low frequency limit of the oscilloscope is reached. The measured
frequency is averaged over $7$ to $8$ periods,  except just before the contact, where no averaging is
applied in order to capture the rapid velocity variations occurring in that
region.

\begin{figure}[tbp]
\begin{center}
{\includegraphics[width=8 cm]{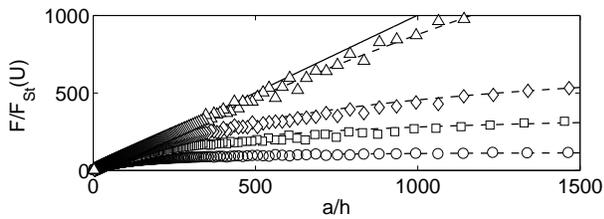}}
\end{center}
\caption{Drag (squares) scaled by the Stokes force for walls decorated with
grooves of similar heights ($e=42-45~\protect\mu$m), but different
$\protect\phi$ and $L$. From top to bottom the data sets for samples \#6, \#2,
\#5, and \#7 (see Table~1). Solid line shows the theoretical prediction for
a smooth lyophilic wall, Eq.(\protect\ref{drag}), defined at the top of
grooves. The dashed curves from top to bottom are calculations of the force
expected for a smooth lyophilic wall shifted from the textured wall to a
distance $s=0.5, 4.2, 9$ and $28.5~\protect\mu$m. }
\label{fig:drag4}
\end{figure}

\textbf{Results and discussion.}-- Fig.~\ref{fig:drag4} shows the drag $F$
(equal to the gravity force) scaled by the Stokes force $F_{St}=6\pi \mu
aU(h)$, i.e. $U(\infty )/U(h)$. Solid line is a theoretical force (Taylor's
equation) predicted for a case of smooth wall and no slippage at the
surface:
\begin{equation}
F_{T}/F_{St} = a/h.  \label{drag}
\end{equation}%
Also included are the experimental data for samples with similar $e$, but
different $\phi $ and $L$, which show deviations from the behavior predicted
by Eq.~\ref{drag}. Close to the wall, for $a/h >50$, the drag is always significantly less than the force near
a smooth wall, and this reduction increases with $\phi $.

\begin{figure}[tbp]
\begin{center}
{\includegraphics[width=8. cm]{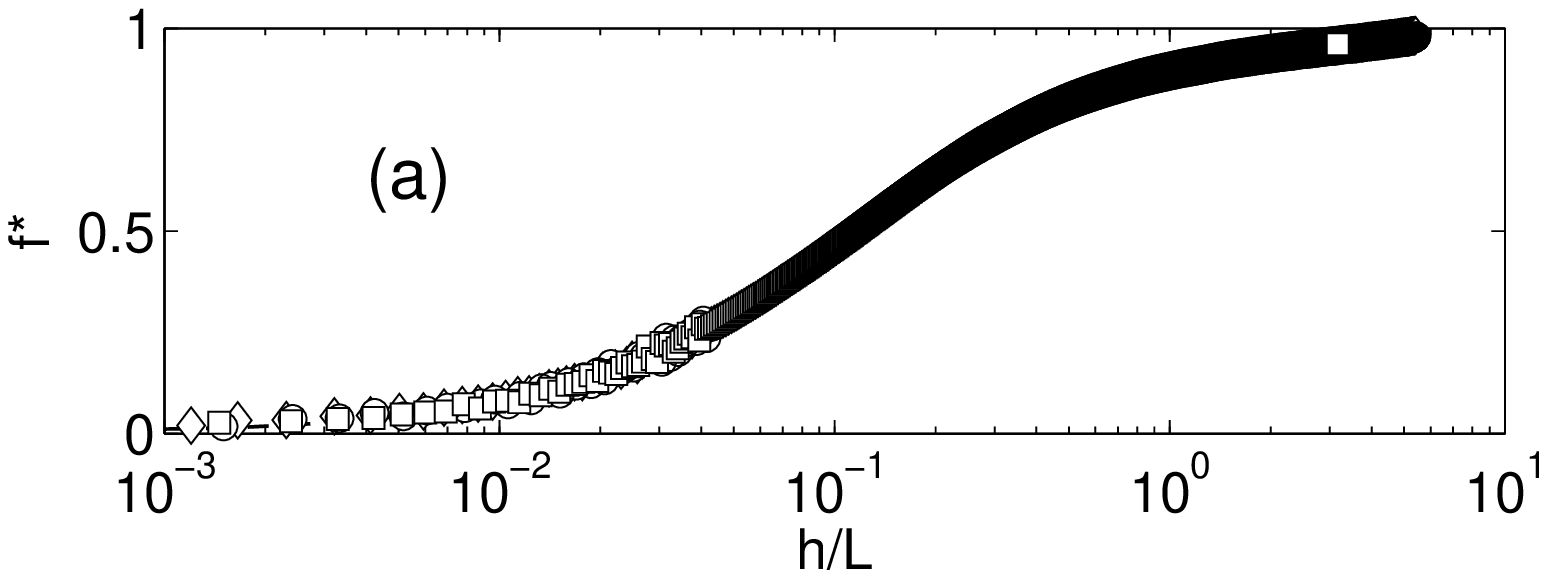}}
{\includegraphics[width=8.
cm]{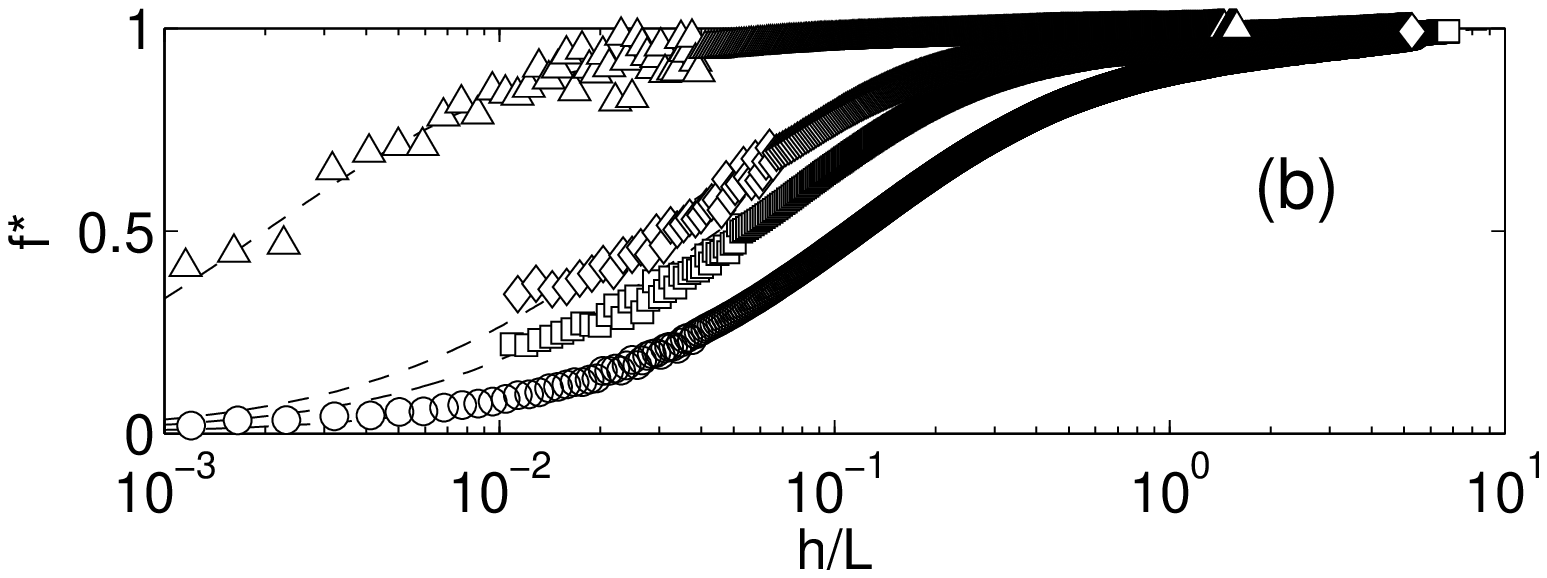}}
\end{center}
\caption{Measured correction to drag (symbols) $\left( a\right) $ for
spheres of different radii ($a=3.5$, $5.75$ and $6.35$~mm) interacting with
sample \#7. Dashed line shows the calculated correction, using
$s=28.5~\protect\mu$m in Eq.(\protect\ref{shift-model}), $\left( b\right) $ for
striped walls having nearly same heights $e=42-45~\protect\mu m$ , but
different $\protect\phi $ and period $L$; from left to right: samples \#6,
2, 5, 7. Dashed lines: model, Eq.\eqref{shift-model}, with (a) $s=28.5~\protect\mu m$, (b) from
left to right: $s=0.5,4.2,9$ and $28.5~\protect\mu m$.}
\label{fig:fstar3radii}
\end{figure}

To examine these deviations we evaluate a correction to the drag force,
\begin{equation}
{f^{\ast }(h)=F(h)/F_{T}(h)}  \label{corr}
\end{equation}%
Note that in general case for a rough surface $f^{\ast }$ should also depend
on the radius of the sphere~\cite{kunert.c:2010}. However, with our
configuration geometry experimental data do not vary with $a$. This is well
illustrated in Fig.~\ref{fig:fstar3radii}(a), where the experimental values
of $f^{\ast }$ obtained with
sample \#7 at different $a$ and plotted as a function of $h/L$
collapse into a single curve, which tends to unity at large distances and
decreases significantly when $h$ becomes of the order of $L$ and smaller. Since at short separations we observe $f^{\ast} \to 0$, one can conclude that slippage (which would lead to $f^{\ast} \to 1/4$~\cite{vinogradova-95}) obviously does not mimic roughness when $h$ is small, by overestimating the drag force. The same remark concerns effective superhydrophobic slippage where $f^{\ast } \to 2 (4-3 \phi)/(8+9 \phi - 9 \phi^2)$~\cite{Asmolov:2011} and  is equal to $\simeq 0.3$ for this particular sample.

Therefore, our experimental results are now compared with theoretical calculations made for an
effective smooth plane shifted down from the top of the corrugations, i.e.
by assuming $F(h)=F_{T}(h+s)$ where $s$ is the value of \emph{constant}, i.e. independent on $h$, shift. This implies that
\begin{equation}
f^{\ast }(h)=\frac{F_{T}(h+s)}{F_{T}(h)}=\frac{h}{h+s}.  \label{shift-model}
\end{equation}%
Fig.~\ref{fig:fstar3radii}(a) includes a calculation  (dashed curve) in which
an adjustable parameter, a  shift of $s=28.5~\mu$m, is incorporated
into the Taylor equation. The fit is very good for all $h$, suggesting the
validity of the model.
Fig.~\ref{fig:fstar3radii}(b) shows another series of experiments made with
the fixed radius of the sphere, but different parameters of the texture. If
similar fits are made to a variety of experiments it is found that the shift
of an equivalent plane required to fit each run increases from 0.5 $\mu $m
for sample \#2 to 28.5 $\mu $m for sample \#7. In Table 1 we present the experimental
values of $s$ for different samples, and theoretical curves calculated with Eq.(\ref{shift-model}) are included in
Figs.~\ref{fig:drag4} and \ref{fig:fstar3radii}. The fit is excellent at all
separations except as very small, $h/L\leq 0.01$. Thus our experiment shows that an effective (scalar) shift, $s$, is a unique physical parameter that fully quantifies drag reduction at a highly anisotropic corrugated surface. This striking result indicates that in our experiment pressure remains isotropic despite an anisotropy of the flow.

Now we try to relate $s$ to parameters of textured surfaces. As
proven in~\cite{Lecoq:2004,Asmolov:2011}, for a large gap, $h\gg L$, the shift of the equivalent no-slip plane from the real
surface is equal to the average of the eigenvalues of the effective
slip-length tensor (at the slip plane defined at the top of asperities)
\begin{equation}
s\simeq \frac{b_{\mathrm{eff}}^{\parallel }+b_{\mathrm{eff}}^{\bot }}{2}.
\label{shift-SHS}
\end{equation}
Therefore, the problem of calculating $s$ reduces to computing the two far-field eigenvalues, $b^{\parallel}_{\rm eff}$ and $b^{\perp}_{\rm eff}$, which
attain the maximal and minimal directional slip lengths, respectively.

\begin{figure}[tbp]
\begin{center}
{\includegraphics[width=8. cm]{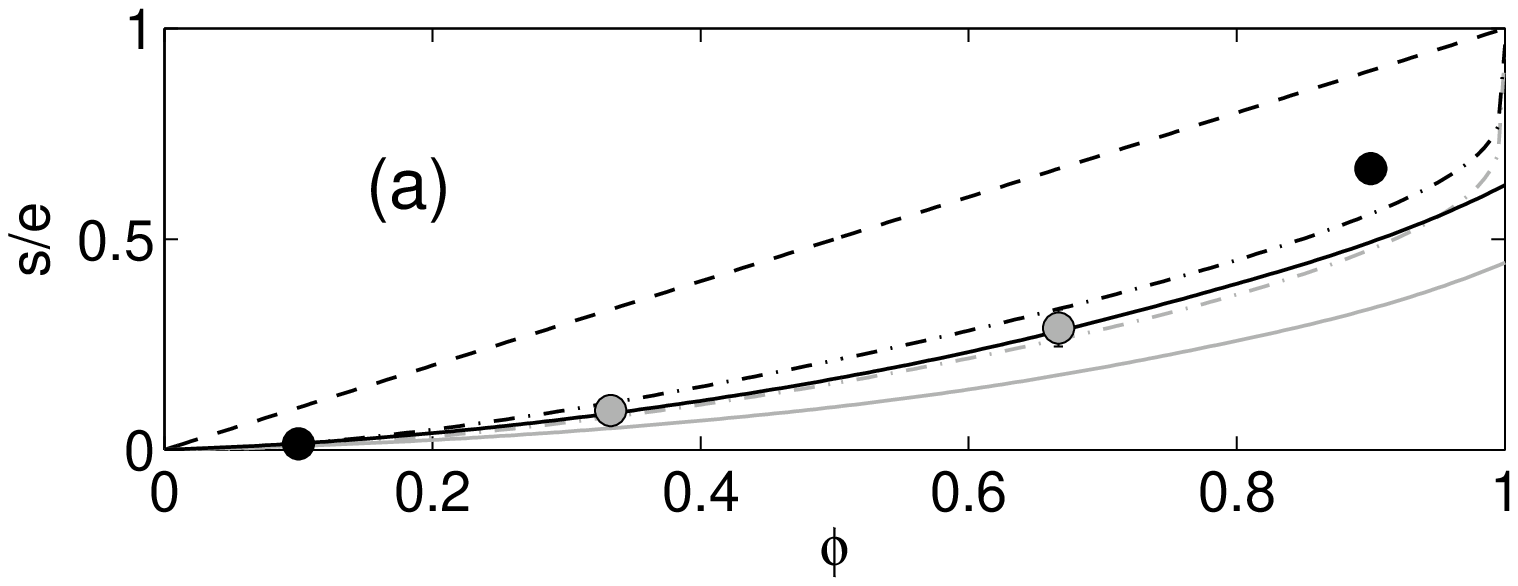}} {\includegraphics[width=8.
cm]{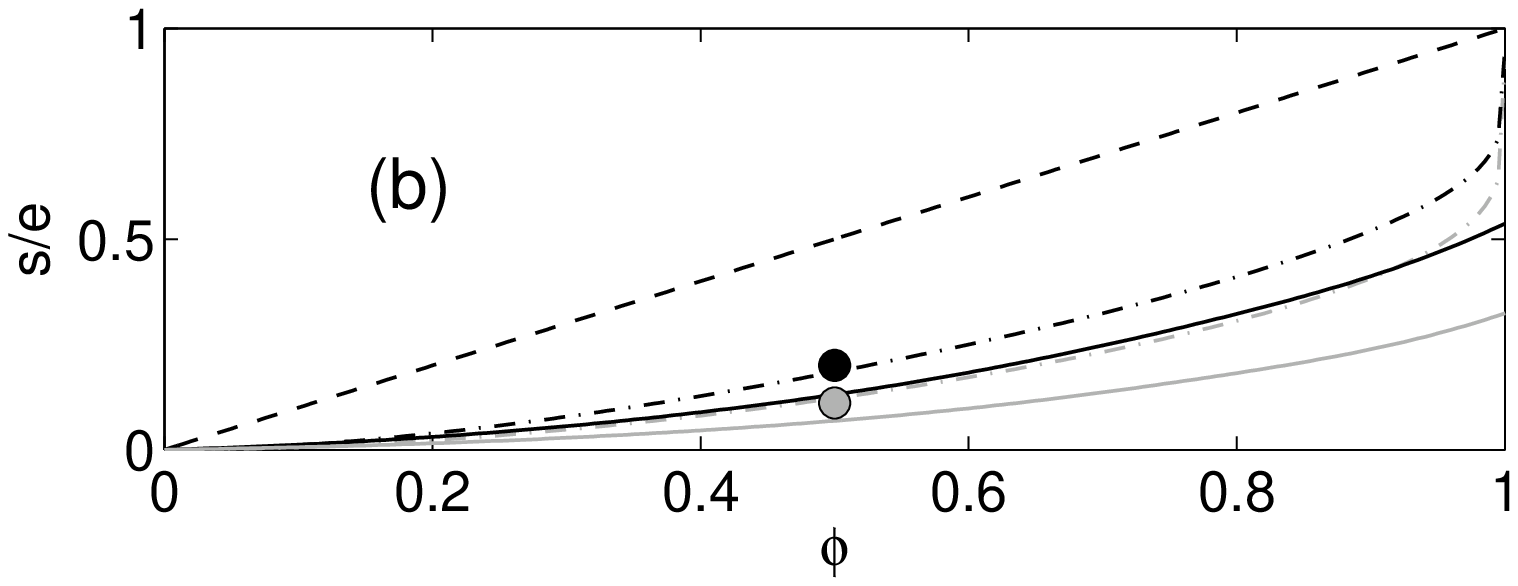}}
\end{center}
\caption{Experimental values of $s$ (symbols) as a function of $\protect\phi $ for
grooves with similar heights ($e=42-45~\protect\mu m$), but different
$\protect\phi$ and $L$. Lines show theoretical predictions, Eq. (\protect\ref{shift-SHS}), where $b_{\mathrm{eff}}^{\parallel}\ $and $b_{\mathrm{eff}}^{\bot }$ are calculated by using the analysis of \protect\cite{wang2003}
(solid lines),  Eqs.~(\ref{b-para})-(\ref{b-perp}) (dash-dotted lines), and Eq.~(\ref{lt})
(dashed lines): (a) samples \#6 and \#7 (black), samples \#2, \#3, (grey);
 (b) sample \#5 (black) and sample \#1 (grey).}
\label{fig:compa1}
\end{figure}

In the limit $e\ll L\ll h,$ the theory \cite{Kamrin_etal:2010}
predicts that the effective no-slip surface for arbitrary smooth periodic
surfaces is at the average height:
\begin{equation}
b_{\mathrm{eff}}^{\parallel, \bot }\simeq\phi e,  \label{lt}
\end{equation}%
so that $s/e$ is controlled
mainly by $\phi $. To examine the significance of $\phi$ more
closely, the experimental $s$ normalized by $e$ are  reproduced in Fig.~\ref{fig:compa1}. The measured data show much smaller $s/e$ than the theoretical prediction of model (\ref{lt}) shown by a dashed line. A possible explanation for this discrepancy
is that the height of asperities in our experiments was not small enough, $0.168\leq e/L\leq 0.45.$ We also compared our data with another calculation (solid curves) for hydrophilic grooves with finite $e/L$ based on numerical results~\cite{wang2003} for eigenvalues of the slip-length-tensor. Even at moderate $e/L$ theoretical predictions for $s$ \cite{wang2003} are much smaller than measured values.

An alternative model can be obtained if we use the
analytic solutions for alternating slip and no-slip stripes~\cite{Belyaev:2010}:
 \begin{equation}
b_{\mathrm{eff}}^{||}\simeq \frac{L}{\pi }\frac{\ln \left[ \sec \left(
\frac{\pi \phi }{2}\right) \right] }{1+\frac{L}{\pi e}\ln \left[ \sec \left(
\frac{\pi \phi }{2}\right) +\tan \left( \frac{\pi \phi }{2}\right) \right] },
\label{b-para}
\end{equation}
\begin{equation}
b_{\mathrm{eff}}^{\bot }\simeq \frac{L}{2\pi }\frac{\ln \left[ \sec \left(
\frac{\pi \phi }{2}\right) \right] }{1+\frac{L}{2\pi e}\ln \left[ \sec
\left( \frac{\pi \phi }{2}\right) +\tan \left( \frac{\pi \phi }{2}\right)
\right] }  \label{b-perp}
\end{equation}%
 where we naturally assumed that the local partial slip is equal to the height of grooves. Fig.~\ref{fig:compa1} shows that Eq. \eqref{shift-SHS}  together with Eqs.~\eqref{b-para}
and \eqref{b-perp} (dashed curves) give almost quantitative agreement with experimental data. (We also include theoretical values of $s$ to Table 1 to allow a direct comparison with experimental results.) Therefore, by using the equivalence of a flow past rough and heterogeneous surfaces at large scale, we were able to quantify a drag reduction at the smaller scale, of the order of the size of roughness elements.
Note however that our results do not apply
to a very thin gap situation $h \ll L,$  where $s$ scales with the channel width~\cite{epaps}, which is again consistent with our experiment.

\textbf{Concluding remarks.}-- We have studied a drag force on a sphere approaching a corrugated plane. Our experiment shows quantitatively that in this situation the effective no-slip boundary condition, which is applied at the imaginary smooth homogeneous isotropic surface located at an intermediate position between top and bottom of grooves, fully mimics the actual one along the true corrugated interface, except as for a very thin gap. The location of this effective isotropic plane depends on the parameters of the texture, and can be found by using simple formulae for effective slip lengths in the limit of a thick channel. Since for grooves anisotropy is maximized, the same conclusion would be valid for other types of anisotropic (e.g., sinusoidal, trapezoidal, and more) and/or isotropic (e.g. pillars, etc) textures, but of course, Eqs.~\eqref{b-para} and \eqref{b-perp} should be replaced by analytical or numerical solutions for a corresponding  texture, as will be described in subsequent papers.

We have also demonstrated that topographically patterned (Wenzel) surfaces could reduce a drag force more efficiently compared to expected even for slipping superhydrophobic (Cassie) textures with trapped gas. Therefore, our results suggested rules and a general strategy for the rational design of topographically patterned surfaces to generate desired low drag.


\bibliography{rough}

\end{document}